# A Network Theory Investigation into the Altered Resting State Functional Connectivity in Attention-Deficit Hyperactivity Disorder

Sadi Md. Redwan[1], Md Palash Uddin[2,3], Muhammad Imran Sharif[4] and Anwaar Ulhaq[5*]

[1]Department of Computer Science and Engineering, University of Rajshahi, Rajshahi 6205, Bangladesh (e-mail: sadi.redwan@ru.ac.bd)

[2]Department of Computer Science and Engineering, Hajee Mohammad Danesh Science and Technology University, Dinajpur 5200, Bangladesh (e-mail: palash_cse@hstu.ac.bd)

[3]School of Information Technology, Deakin University, Geelong, VIC 3220, Australia

[4]COMSATS University Islamabad, Wah Campus, Punjab 47040, Pakistan (e-mail: mimraansharif@gmail.com)

[5]School of Computing. Mathematics and Engineering, Charles Sturt University, NSW, Australia (email: aulhaq@csu.edu.au)

[*]Corresponding author

**Abstract**

In the last two decades, functional magnetic resonance imaging (fMRI) has emerged as one of the most effective technologies in clinical research of the human brain. fMRI allows researchers to study healthy and pathological brains while they perform various neuropsychological functions. Beyond task-related activations, the human brain has some intrinsic activity at a task-negative (resting) state that surprisingly consumes a lot of energy to support communication among neurons. Recent neuroimaging research has also seen an increase in modelling and analysing brain activity in terms of a graph or network. Since graph models facilitate a systems-theoretic explanation of the brain, they have become increasingly relevant with advances in network science and the popularization of complex systems theory. The purpose of this study is to look into the abnormalities in resting brain functions in adults with Attention Deficit Hyperactivity Disorder (ADHD). The primary goal is to investigate resting-state functional connectivity (FC), which can be construed as a significant temporal coincidence in blood-oxygen-level dependent (BOLD) signals between functionally related brain regions in the absence of any stimulus or task. When compared to healthy controls, ADHD patients have lower average connectivity in the Supramarginal Gyrus and Superior Parietal Lobule, but higher connectivity in the Lateral Occipital Cortex and Inferior Temporal Gyrus. We also hypothesise that the network organisation of several default mode and dorsal attention regions is abnormal in ADHD patients.

## 1. Introduction

Attention-Deficit Hyperactivity Disorder (ADHD) is a neurological and mental disorder marked by inattention, hyperactivity, excessive physical activity, and impulsivity. It occurs early in life. In 60% of the cases, it lasts into adulthood, and the symptoms range from mild to severe in adult ADHD patients [1]. Diagnosis of ADHD still involves classical pathological methods that mostly rely on cognitive and behavioral tests. The specific characteristics of its underlying atypical brain development are still difficult to synthesize into a coherent account, despite the fact that numerous studies indicate distinct characteristics of the pathological brain that can be associated with neurodevelopmental dysfunction and identified using modern electrogram and neuroimaging techniques [2].

Human cognition has been linked to resting state functional connectivity (FC) patterns, and the salience network and the Default Mode Network (DMN) have been investigated as resting state networks concerning cognitive activity [3][4]. The integration of cognitive and emotional processing, as well as environmental monitoring, has been linked to the DMN [3][4]. Furthermore, increased DMN activity has been linked to an increase in the occurrence of stimulus-independent thoughts [5][6]. A recent EEG study linked abnormal DMN organisation to Social Anxiety Disorder [7]. With a negative association between the connectivity of the DMN and the degree of neurodivergence, the potential efficacy of RS-fMRI in the diagnosis of neurobiological and neurocognitive disorders has also been established [8]. Patients with Major Depressive Disorder (MDD) [9][35] and Schizophrenia [10][11] have been successfully distinguished from healthy controls using RS-fMRI pattern categorization. The atypical brain functions of patients with Autism Spectrum Disorder (ASD) have also been studied using RS-fMRI recently [12]. The domain of RS-fMRI studies is not limited to explanations of neuropsychiatric disorders that are already known to be linked with atypical brain development, it also encompasses the effects of other phenomena. The success of RS-fMRI has recently been translated into analyzing the effects of Problematic Internet Use (PIU) in another study [13]. RS-fMRI has also been used to investigate functional Dyspepsia [34] and obesity [36].

The main focus of most RS-fMRI studies has been functional connectivity (FC). The brain has several anatomical regions that are functionally connected such that fluctuations in the BOLD signals in these regions show significant correlation during the performance of a task or during the resting state. Analyzing the correlation between two regions enables us to comprehensively deduce the neuronal activity between them. This correlation can also be explained as a pairwise relationship between two regions, and ultimately we reach a graph model of the brain with dynamic temporal activity among different anatomical regions. Evidently, modeling and analyzing the brain as a network or graph has become mainstream in neuroimaging studies in recent years. The brain's graph model is a generic representation of the brain's connections, representing either functional connectivity among neural ensembles or white-matter fiber tracts' structural relationships [14]. Graph-based models have already been successful in identifying dissimilarities in the functional organization of ASD [15]. Even though there is ample proof of functional brain network anomalies in ASD, graph models have provided new insights leveraging the large dataset Autism Brain Imaging Data Exchange (ABIDE) [16]. Graph models have also been used in analysing patients with Hemianopia [37], schizophrenia [38], and brain metastases [39]. Taking these findings into account, we focus on utilizing network theory and graph-based models to investigate the functional connectivity difference between healthy controls and ADHD patients using the Consortium for Neuropsychiatric Phenomics (CNP) dataset [17]. We aim to show how the altered resting-state functional connectivity in adult ADHD differs from the healthy controls based on several methods, including random field theory, independent component analysis (ICA), and network-based statistic (NBS).

## 2. Materials and Methods

### 2.1 Resting-State fMRI (RS-fMRI)

Functional magnetic resonance imaging (fMRI) has expanded in popularity, sophistication, and range of applications since its inception more than two decades ago. fMRI uses blood-oxygen-level dependent (BOLD) contrast imaging to measure the brain's neuronal activity based on cerebral blood flow variations – which has proven remarkably useful in determining which parts of the brain are most active. The application of fMRI in stimulus-based paradigms has been crucial to our current conception of how the brain works in task-specific situations. A lot of contemporary research is based on presumptions of which regions of interest (ROI) of the brain are active when a certain activity is being performed. This entails tracking the relative variations in BOLD signal from baseline in various brain areas while the tasks are being performed or in the presence of certain stimuli [18]. Even in the absence of any stimulus, BOLD signals show some spontaneous fluctuations in certain cortical and subcortical regions. In general, an increasing corpus of neuroimaging studies maintains that intrinsic neural activity is at least partially responsible for resting-state BOLD fluctuations [19]. Subsequently, resting-state fMRI (RS-fMRI) has been popularized for brain mapping to conduct studies on spontaneous regional interactions that take place in the absence of any explicit task or when the brain is at rest.

The UCLA Neuropsychiatric Phenomics Consortium released a dataset in 2016 containing neuroimaging and phenotypic information for 272 individuals. It includes 130 healthy controls, 43 patients with adult ADHD, 49 with bipolar disorder, and 50 with schizophrenia. The dataset includes T1 weighted anatomical images and fMRI scans of the subjects during various task protocols, including resting state (eyes open). Gorgolewski et al. used FreeSurfer and fMRIPrep pipelines to preprocess the dataset, which was published in 2017 [20]. Since the Los Angeles County Department of Mental Health and the Institutional Review Board at UCLA approved their procedures, preprocessing the entire dataset would be redundant for this work. For our study, we preprocess RS-fMRI data of 40 ADHD patients and 40 random healthy controls based on the availability of both anatomical and functional data in the original dataset. The demographic information of the selected subjects is shown in Table 1.

Table 1. Subject demographic information

|                   | Control      | ADHD         |
| ----------------- | ------------ | ------------ |
| N (Male, Female)  | 40 (20, 20)  | 40 (21, 20)  |
| Mean Age (SD)     | 29.93 (8.59) | 32.05 (10.41)|

We obtain the resting state data from OpenfMRI [21] (accession number: ds000030). The Public Domain Dedication and License v1.0 allow the dataset to be used and redistributed without restriction. The data descriptor adheres to the ODC Attribution-Sharealike Community Norms [17].

## 2.2 Preprocessing

The resting fMRI scan lasted 304s, according to the data descriptor. During the scan, the participants were not given any stimuli and were told to remain calm while keeping their eyes open. The fMRI data were acquired using an echoplanar imaging (EPI) sequence, which had the following specifications: slice thickness of 4 mm, 34 slices, TR of 2s, TE of 30ms, flip angle of 90°, matrix of 64×64, FOV of 192 mm, and oblique slice orientation [17].

We preprocess the individual subject's data using SPM12 (statistical parametric mapping). After the standard realignment and slice-time correction procedure, we use ART (Artifact Detection Tools) for outlier detection. For normalization, we choose to apply direct normalization to the MNI152 space. Finally, after structural segmentation and normalization, a Gaussian kernel of 6mm FWHM is applied with the aim of increasing the BOLD signal-to-noise ratio.

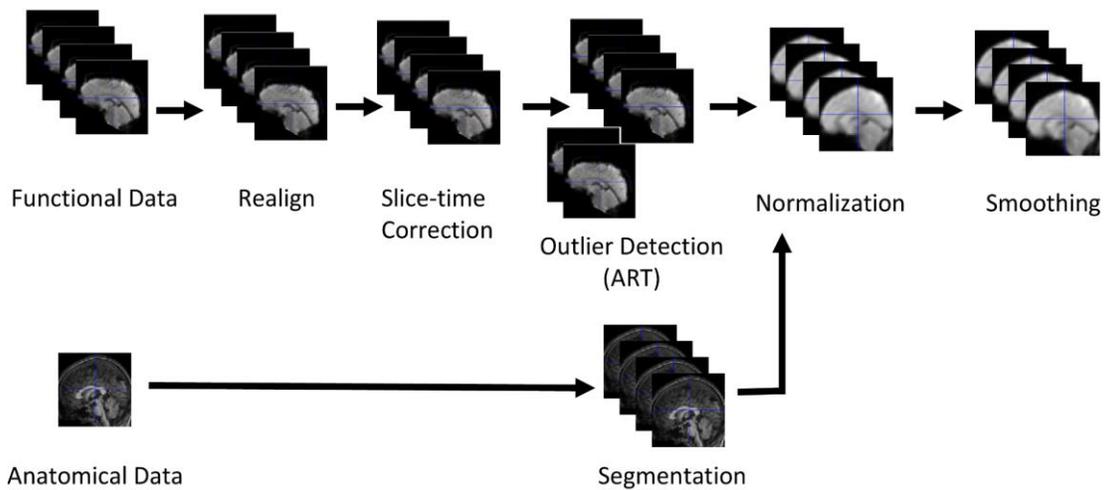

Figure 1. SPM12 preprocessing pipeline. Functional data is realigned by registering to the average 3D image. Slice-time is corrected by using the median slice as a reference. Potential outliers with displacement above a threshold (0.9mm) are identified, and SPM12 segmentation and normalization are applied. Smoothing is performed using a Gaussian kernel as the final step.

Before further processing, BOLD signals are filtered in time domain using a band-pass filter of 0.008 to 0.09 Hz to focus on low-frequency oscillations while reducing noise from noise sources, including physiological, head-motion, and other noise components. Ordinary least squares (OLS) regression is used to independently eliminate several nuisance regressors during denoising. These include estimated subject-motion parameters [23], session and task effects [24], and white matter and cerebrospinal region noise components [22]. In order to evaluate the procedure's impact, we estimate the distribution of FC values between randomly selected pairs of ROI(s) in the brain before and after denoising. Figure 2 displays the distributions.

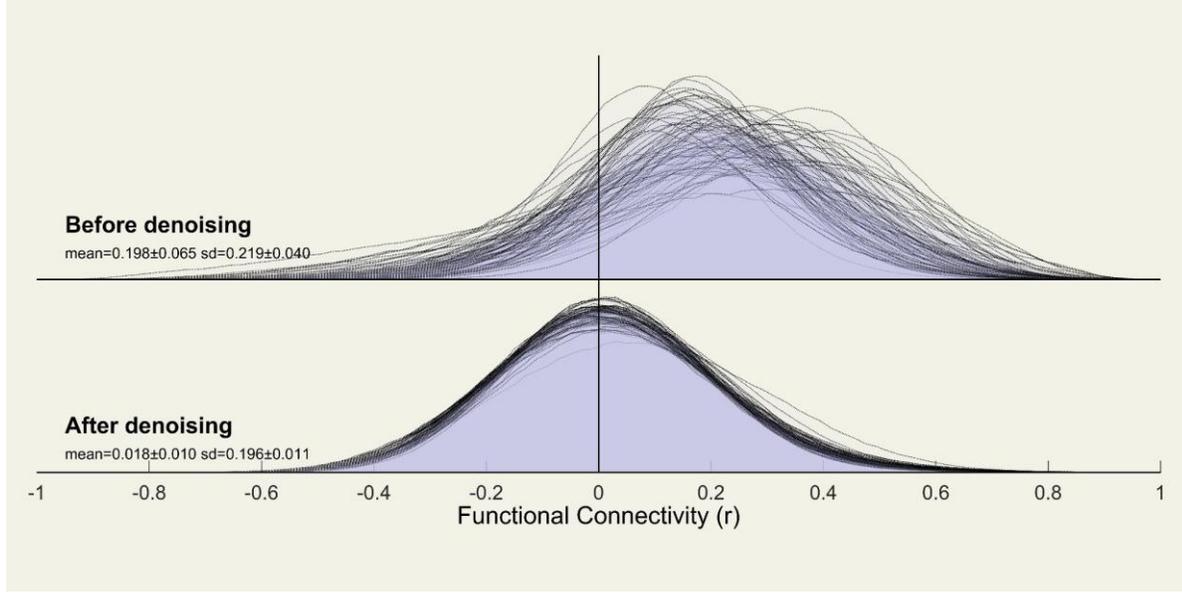

Figure 2. Effect of denoising. Distributions of random FC show excessive variability and bias before denoising (top). Denoising reduces the variability significantly, producing centred distributions (bottom).

**2.3 First-Level Analysis**

First-level analysis refers to the process of modeling each subject's time-series data using a general linear model (GLM). There are multiple ways to segment the human brain into different regions of interest (ROI), such as functional regions based on a meta-analysis [25] and spatially constrained spectral clustering [26]. We employ anatomical ROI based on the Harvard-Oxford cortical and subcortical atlas and the AAL cerebellar atlas, which is the default atlas included with the CONN toolbox.

The atlas ROI(s) that divide the brain into different regions are also referred to as seeds. These ROIs were used to build a GLM to best fit the data observed in each voxel during the denoising step. After that, a correlation analysis is performed, averaging across the signal in each ROI(s) and correlating it with other ROI(s) in the brain (ROI-to-ROI), as well as generating correlation maps with each voxel as a seed (Seed-to-Voxel). With a predetermined seed or ROI, seed-based connectivity (SBC) metrics describe the connection patterns and show the degree of functional connectivity between each voxel or region in the brain and the seed or ROI. These are calculated as the bivariate correlation coefficients between an ROI BOLD time-series and each distinct voxel BOLD time-series using the Fisher transform defined as follows.

$$r(x) = \frac{\int S(x,t)R(t)dt}{\left(\int R^2(t)dt \int S^2(x,t)dt\right)^{1/2}}$$

$$Z(x) = \tanh^{-1}(r(x))$$

Here, $S$ is the BOLD time-series, $r$ is the map of correlation coefficients, and $Z$ is the Fisher-transformed SBC map.

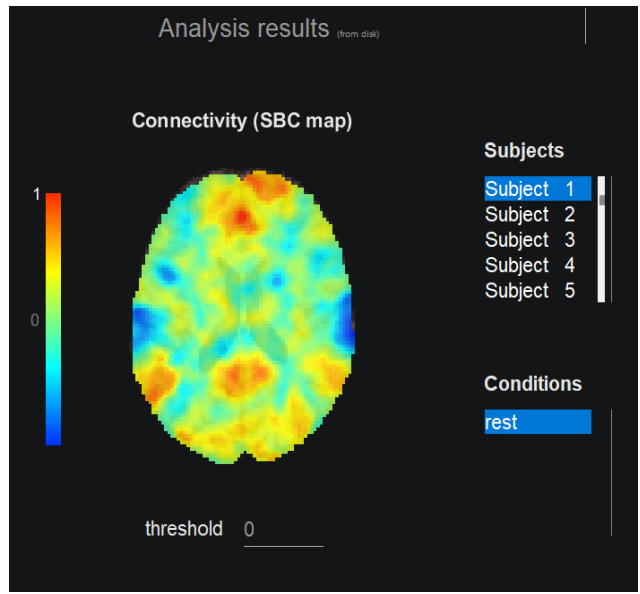

Figure 3. SBC map (single subject). The heatmap shows which ROI(s) have more functional connectivity with all the other voxels. This is effectively a map of Fisher-transformed bivariate correlation coefficients between the ROI and voxel signals.

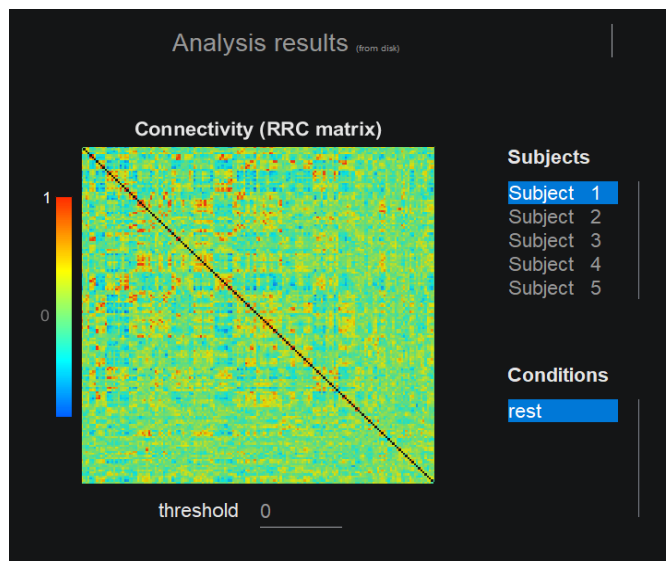

Figure 4. RRC matrix (single subject). This is effectively an adjacency matrix of the ROI(s) with the same correlation measures as SBC. A higher correlation between the BOLD signals in a pair of ROI(s) indicates increased functional connectivity between them.

In a predefined set of regions, ROI-to-ROI connectivity (RRC) metrics describe the connectedness between all pairs of ROI(s). The formulation of these measurements avoids the SBC asymmetry between ROI(s) and targets (voxels) while adhering to the exact same organization and features as the SBC measures.

## 2.4 Group-Level Analysis

In this stage, the outcomes of the first-level analyses of each subject are merged for an analysis of the entire population. The effect estimates (GLM) are often integrated across subjects in group-level analysis using the t-test, ANOVA, ANCOVA, multiple regression, or linear mixed-effects (LME) models [27]. The GLM framework only requires the specification of four parameters, which are associated with subject-effects (*X*), conditions (*Y*), between-subjects contrast (*C*), and between-conditions contrast (*M*) matrices. Thus, a very broad range of classical analyses can be specified using the same GLM framework. Different analyses specified by different parameters in this framework are shown in Table 2. We chose between-subjects contrast [1, -1] for a two-sample t-test. Effects of each group (regression) were also computed for further ROI analyses in the future.

Table 2. Group-level analysis parameters. One-sample t-test measures the FC difference from a specific value. Regression measures whether the FC of a certain group differs from zero (the entire population). Two-sample t-test measures the FC difference between two groups.

| Analysis type | Subject-effects (*X*) | Between subjects contrast (*C*) | Conditions (*Y*) | Between conditions contrast (*M*) |
|---|---|---|---|---|
| one-sample t-test | All Subjects | [1] | Rest | [1] |
| two-sample t-test | Patients, Controls | [1, -1] | Rest | [1] |
| regression | All Subjects, Behavioral | [0, 1] | Rest | [1] |
| one-way ANCOVA covariate control | Patients, Controls, Age | [1, -1, 0] | Rest | [1] |

## 2.5 Functional Connectivity (FC)

For FC analysis, we employ seed-based functional connectivity methodology based on random field theory [28] and functional network connectivity (FNC) technique based on spatial independent component analysis (ICA) [29] and network-based statistic (NBS) [14]. By explicitly examining correlations, the random field theory approach establishes a critical value at which the correlations are meaningful. Most recent FC studies have adopted the computationally efficient blind source separation (BSS) method known as ICA [30][31]. We use an ICA implementation based on the group-ICA methodology with optional subject-level dimensionality reduction, subject concatenation, and the fastICA algorithm for group-level independent component formation [32][33]. The cluster-level inference used in the FNC approach is based on multivariate statistics by considering groups or networks of related ROI(s). A standard group-level GLM analysis of RRC matrices produces a single statistical matrix of *F*-values, characterizing the effect of interest. NBS, in comparison, begins with a two-dimensional statistical parametric map made up of the whole ROI-to-ROI matrix of *T*-statistics or *F*-statistics computed using the GLM. The network-level parameters of our FNC and NBS analysis are discussed in Table 5 and Table 6 respectively. The experiments were done using MATLAB R2022b on Microsoft Windows 11 (22H2) platform and FSL6.0 in Ubuntu 22.04 platform on an AMD Ryzen 7 3750H computer.

## 3. Result and Discussion

### 3.1 Random Field Theory

The random field theory (RFT) based inference used in this work uses a statistical parametric map of *T*-values estimated using the GLM. This map is first thresholded using a priori threshold levels (voxel threshold p<0.01, cluster threshold p<0.01). Suprathreshold regions that are derived as an outcome identify a number of non-overlapping clusters. The following ROI(s) are identified in this analysis to have decreased average connectivity in ADHD patients compared to the healthy controls. The regions are visualized in Figure 5 and Figure 6.

Table 3. Average connectivity difference (CONTROL>ADHD). The ROI(s) are sorted by cluster size (number of voxels). Area denotes the percentage of the ROI covered by the voxels.

| ROI | Hemisphere | Voxels | Area |
|---|---|---|---|
| Supramarginal Gyrus, posterior division | Right | 131 | 11% |
| Supramarginal Gyrus, anterior division | Left | 98 | 10% |
| Superior Parietal Lobule | Right | 91 | 6% |
| Superior Parietal Lobule | Left | 71 | 5% |

The following ROI(s) show increased average connectivity in ADHD patients compared to the healthy controls, as shown in Figure 5 and Figure 6.

Table 4. Average connectivity difference (ADHD>CONTROL).

| ROI | Hemisphere | Voxels | Area |
|---|---|---|---|
| Lateral Occipital Cortex, superior division | Right | 99 | 2% |
| Inferior Temporal Gyrus, anterior division | Left | 84 | 25% |
| Inferior Temporal Gyrus, posterior division | Left | 36 | 4% |

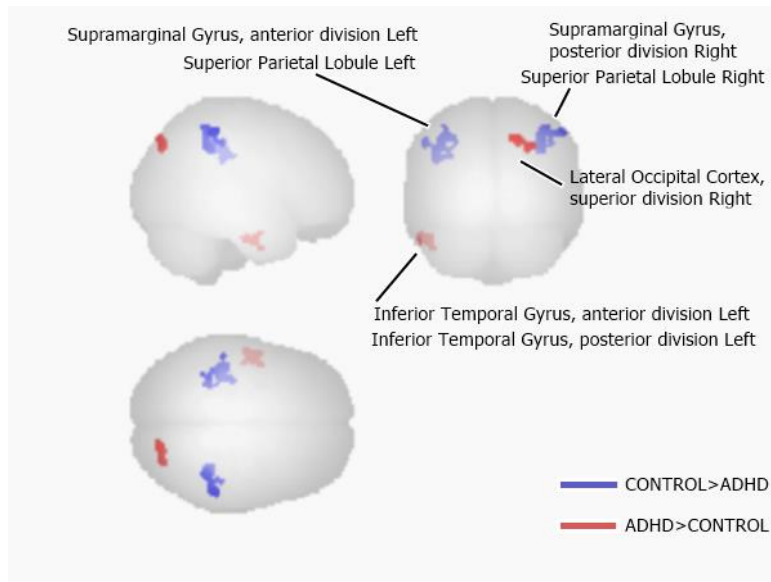

Figure 5. ROI(s) with the most significant average connectivity difference. ROI(s) where the average connectivity is higher in the control population are highlighted in blue, and ROI(s) where the average connectivity is higher in the ADHD population are highlighted in red. Cluster-size *p*-FDR corrected with threshold *p*<0.01

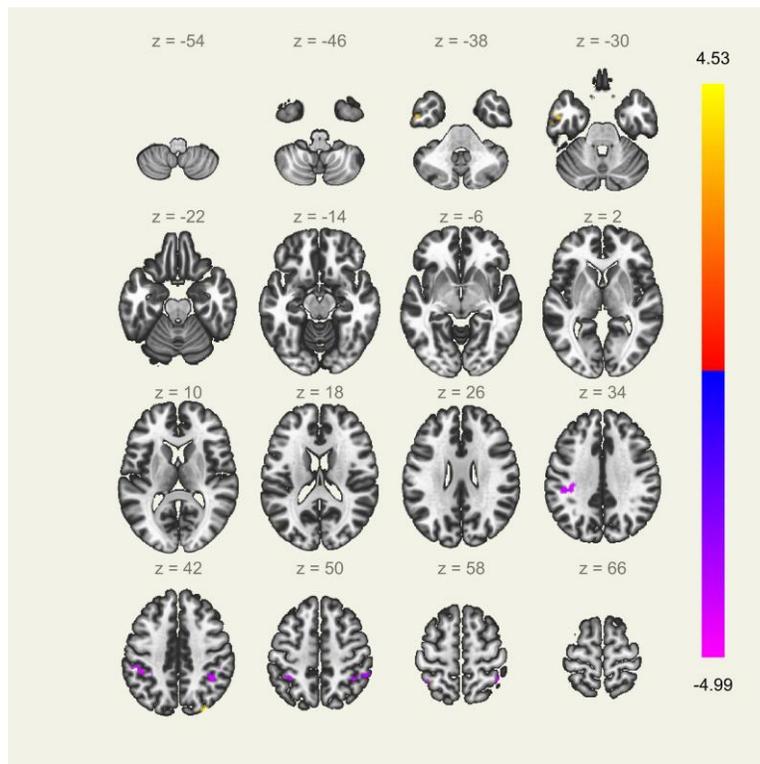

Figure 6. Random field theory parametric statistical inference is shown in the axial plane with the clusters highlighted. Lower average connectivity and higher average connectivity in the ADHD population are denoted by lower and higher values, respectively.

## 3.2 Functional Network Connectivity

To compute an *F*-statistic for each pair of networks and a *T*-statistic for each connection, FNC evaluates the complete connections among all ROIs in the intra-network and inter-network connectivity sets. Then, using an FDR-corrected cluster-level *p*-value ($p<0.1$), which is defined as the anticipated percentage of false discoveries among all pairs of networks, the largest network clusters with significant differences are determined (MVPA omnibus test). Table 5 discusses the statistics at the network level.

Table 5. FNC statistics (connection threshold: $p<0.1$, cluster threshold $p<0.1$)

| Analysis unit | Statistic | *p*-uncorrected | *p*-FDR |
| --- | --- | --- | --- |
| Cluster 1/276 | F(3,76) = 6.92 | 0.000352 | 0.083580 |
| Cluster 2/276 | F(3,76) = 6.44 | 0.000606 | 0.083580 |

In ADHD, both the Medial Prefrontal Cortex (MPFC) and the Medial Frontal Cortex (MedFC) have increased functional connectivity with the Middle Temporal Gyrus, anterior and posterior divisions (aMTG, pMTG), and the Inferior Temporal Gyrus, anterior and posterior divisions (aITG, pITG). Figure 7 shows that the MPFC has decreased connectivity with the Intraparietal Sulcus (IPS) and Superior Parietal Lobule (SPL). MPFC and IPS are notably known as a major DMN region and a dorsal attention region respectively.

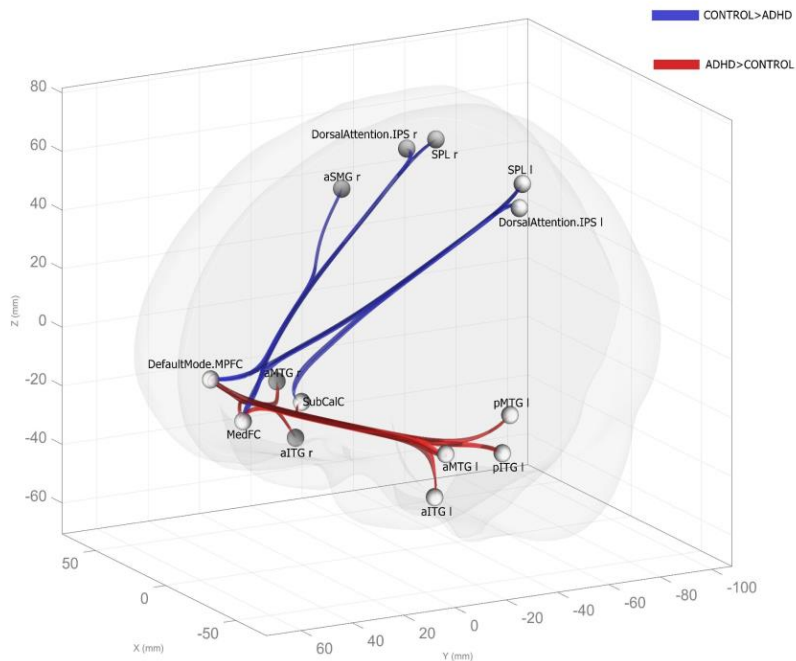

Figure 7. FNC summary - connections drawn in blue denote higher functional connectivity between ROI(s) in the control population and connections drawn in red denote higher functional connectivity between ROI(s) in the ADHD patients.

### 3.3 Network-Based Statistic

NBS is used in this work by applying connection threshold p<0.2 and cluster threshold p<0.05, resulting in suprathreshold connections that define a graph encompassing the whole brain, thus making an inference of the entire network of ROI(s). This graph is then fractionated into connected subnetworks. The network mass (sum of *F*-squared statistics) over all connections within the subnetworks that are identified to show significant dissimilarity between the two groups are discussed in Table 6.

Table 6. NBS (connection threshold: $p<0.2$, cluster threshold $p<0.05$)

| Analysis unit | Statistic | *p*-uncorrected | *p*-FDR | *p*-FWE |
|---|---|---|---|---|
| Network 1/5 | Mass = 182.95 | 0.003823 | 0.019116 | 0.013000 |
| Network 2/5 | Mass = 102.02 | 0.012611 | 0.031526 | 0.032000 |

NBS analysis shows increased functional connectivity between Medial Prefrontal Cortex (MPFC) and Inferior Temporal Gyrus, anterior division (aITG) in ADHD. Similar to the FNC analysis, Intraparietal Sulcus (IPS) shows decreased connectivity with MPFC in ADHD patients compared to the controls. Interestingly, the Paracingulate Gyrus (PaCiG r) also shows decreased connectivity with IPS and Heschl's Gyrus (HG), as shown in Figure 8.

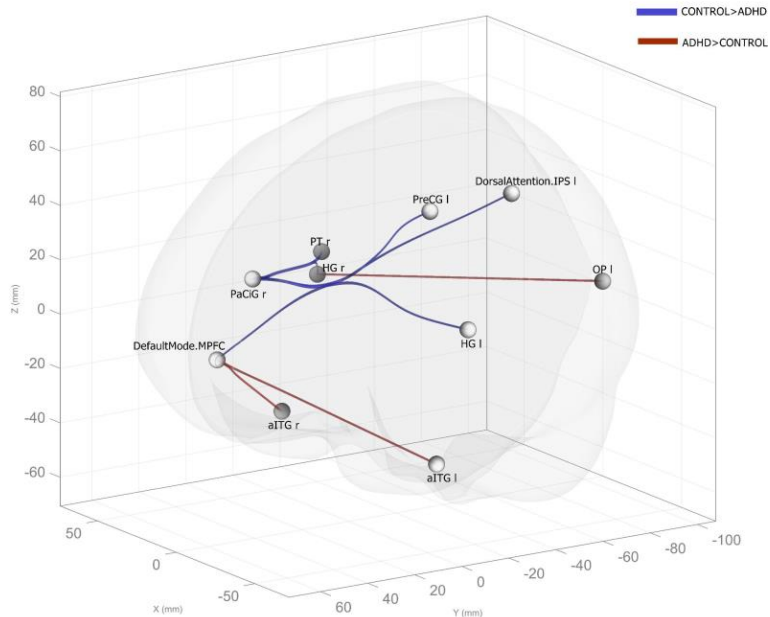

Figure 8. NBS summary - connections drawn in blue denote higher functional connectivity between ROI(s) in the control population and connections drawn in red denote higher functional connectivity between ROI(s) in the ADHD patients.

## 4. Conclusion

In this investigation, we used the UCLA-CNP dataset to preprocess, denoise, and analyse RS-fMRI data from healthy adults and ADHD patients. We have demonstrated the potential utility of network theory in identifying the brain regions and regional interactions that result in non-typical resting-state brain activity in ADHD. In ADHD patients, global average connectivity in the Supramarginal Gyrus and Superior Parietal Lobule is lower compared to healthy controls, with higher average connectivity in the Lateral Occipital Cortex and Inferior Temporal Gyrus. In multiple analyses, one of the major DMN regions, the Medial Prefrontal Cortex, showed increased functional connectivity with the Inferior Temporal Gyrus and decreased functional connectivity with the Intraparietal Sulcus in ADHD patients. We hope that these findings will facilitate further investigation to understand brain connectivity in patients with Attention-Deficit Hyperactivity Disorder.